# Interface enhanced electron-phonon coupling and high temperature superconductivity in potassium-coated ultra-thin FeSe films on SrTiO$_3$

Chenjia Tang,[1] Chong Liu,[1] Guanyu Zhou,[1] Fangsen Li,[1] Ding Zhang,[1] Zheng Li,[1] Canli Song,[1,2] Shuaihua Ji,[1,2] Ke He,[1,2] Xi Chen,[1,2] Lili Wang,[1,2,*] Xucun Ma,[1,2] and Qi-Kun Xue[1,2,†]

[1] *State Key Laboratory of Low-Dimensional Quantum Physics, Department of Physics, Tsinghua University, Beijing 100084, People's Republic of China*

[2] *Collaborative Innovation Center of Quantum Matter, Beijing 100084, People's Republic of China*

Alkali-metal (potassium) adsorption on FeSe thin films with thickness from two unit cells (UC) to 4-UC on SrTiO$_3$ grown by molecular beam epitaxy is investigated with a low-temperature scanning tunneling microscope. At appropriate potassium coverage (0.2−0.3 monolayer), the tunneling spectra of the films all exhibit a superconducting-like gap larger than 11 meV (five times the gap value of bulk FeSe), and two distinct features of characteristic phonon modes at ~11 meV and ~21 meV. The results reveal the critical role of the interface enhanced electron-phonon coupling for possible high temperature superconductivity in the system and is consistent with recent theories. Our study provides compelling evidence for the conventional pairing mechanism for this type of heterostructure superconducting systems.

Recent report on the high $T_C$ superconductivity in the heterostructure of single unit-cell (UC) FeSe films on SrTiO$_3$ (STO) (001) substrates grown by molecular beam epitaxy (MBE) [1] has stimulated considerable research interests in superconductivity community. The FeSe/STO system displays superconducting gaps Δ ~15−20 meV [2-6] and $T_C$ above 65 K [7-10], almost one order of magnitude higher than the values (Δ ~2.2 meV and $T_C$ ~8 K) of bulk FeSe [11, 12]. Interestingly, a unique Fermi surface topology is found in FeSe/STO: only electron-like pockets exist at the Brillouin zone corners and no hole pockets in the Brillouin zone center [2-5], which has experimentally been demonstrated to be induced by charge transfer from the oxygen vacancies in the STO substrates to 1-UC FeSe films above [2-5, 13]. On the other hand, with hole pockets appearing in the Brillouin zone center and becoming stronger with increasing thickness [2, 4, 5], multilayer (≥ 2-UC) films on STO prepared by the same method don't exhibit any signature of superconductivity [1, 2, 4-6].

Considering the fact that the superconductivity in cuprate and iron-based layered superconductors is similarly achieved by doping (of a parent Mott insulator or metallic compound) and that the transition temperature $T_C$ could be tuned by the amount of doped carriers in phase diagram [14, 15], one may speculate that the absence of superconductivity in multilayer FeSe films is due to insufficient carrier transfer from STO substrate. This is demonstrated by a recent temperature-dependent angle resolved photoemission spectroscopy (ARPES) study which shows that, once coated with potassium (K) atoms, 3-UC FeSe films become electron populous and exhibit a pairing formation temperature of 48 ± 3 K at optimal doping [16]. By observing a dome-shaped phase diagram, the study suggests that the high $T_C$ probably comes from antiferromagnetic fluctuation that is enhanced at interface and facilitated by forming some ordered phase. However, there is no proof for the parent ordered phase, thereby its link to the antiferromagnetic interaction enhancement or suppression as a function of doping level is speculative.

Similar to our initial proposal [1], the coupling between electrons and phonons that drives the formation of electron pairs as in conventional superconductors has been suggested to be responsible for the high $T_C$ in FeSe/STO, which is evidenced by the ARPES observation of shake-off bands [5]. The experiment suggests the important role of electron-phonon coupling. In conventional superconductors, the electron-phonon (*e-ph*) coupling is characterized by the dips in the second derivative of tunneling conductance ($d^2I/dV^2$) that correspond to the peaks in Eliashberg spectral function $α^2F(ω)$ at energy E = Ω + Δ (Ω is the phonon energy and Δ the superconducting gap) [17-21]. Such features were observed in high temperature superconductors and often interpreted as bosonic modes [22-25]. However, the nature of those bosons and hence the pairing mechanism is under hot debate.

In this Letter, by measuring tunneling conductance ($dI/dV$) of K-coated 2−4 UC FeSe thin films on STO with the scanning tunneling microscopy/spectroscopy (STM/STS) technique, we find that they all become superconducting at appropriate K coverage and exhibit a U-shaped gap larger than 11 meV. Furthermore, similar to conventional superconductors, the emergency of the superconducting gap is always accompanied with characteristic phonon modes, whose frequencies are ~11 meV and ~21 meV for all of the K-coated 2−4 UC FeSe films and bare 1-UC FeSe films. Our experimental finding further demonstrates the pivotal role of *e-ph* coupling in this system and thus supports our initial proposal of the interface enhancement reported in Ref. [1].

The FeSe thin films were grown on STO substrates by MBE, the details of the growth can be found in our previous studies [1, 7]. For a given thickness, systematic K deposition from 0.01 monolayer (ML) to 0.30 ML was performed, and *in situ* STM/STS measurement was then conducted to investigate the morphology and tunneling spectra at each K coverage, as detailed in the Supplementary. Here, 1 ML is defined as the coverage at which K atoms occupy all the hollow sites of Se lattice as in the case of stoichiometric $K_1Fe_2Se_2$ [26]. As shown in Figure 1, below 0.20 ML, K atoms adsorb individually on the surface and occupy the hollow sites of the (1 × 1) Se terminated (001) surface. With increasing coverage, some of K atoms pile up and form clusters. Regardless of the clusters formation, we see no change in the overall film terrace-step morphology even at the maximum coverage, 0.30 ML. Thus, we assume that the structure of underlying FeSe films remains undisturbed, and the major change is in the electron density owing to electron donation from K adatoms (probably also from clusters), which converts the films into superconducting state as discussed below.

Figure 2 summarizes the differential tunneling spectra ($dI/dV$) of 2−4 UC films at various K coverage. We can clearly see that at approximately 0.1 ML (roughly 0.05 electron/Fe), a U-shaped superconducting-like gap develops in all original non-superconducting films. The U-shaped gap with vanishing density of state at $E_F$ is similar to that of 1-UC films on STO [1], primarily indicating node-less pairing. With increasing K coverage, the gap increases and reaches a maximum value of 14.5 ± 1.0 meV, 13.1 ± 1.4 meV and 11.9 ± 1.4 meV for 2-UC, 3-UC and 4-UC films, respectively. The optimal K coverage, defined as the coverage at which the gap reaches maximum, for 2−4 UC films is found similar, 0.20 ML (roughly 0.1 electron/Fe), coincident with the coverage at which significant cluster formation occurs. Further increasing K coverage alters the gap size little within experimental uncertainty. An intuitive explanation for this observation is clustering of K atoms, since clustering consumes electrons available for doping (Fig. S1).

The second important finding of this study is the observation of characteristic bosonic modes. The data for 3-UC film is shown in Figure 3(a). We choose particularly the K-coated 3-UC film for detailed discussion since its superconductivity has been demonstrated by recent ARPES experiment [16]. In addition to the superconducting gap Δ ~11.7 meV in this case, we can see some satellite dip-hump features outside the coherence peaks [top panel, Fig. 3(a)]. These features are more clearly identified in the normalized $dI/dV$ spectrum [middle panel, Fig. 3(a)], and bear striking resemblance to those of phonons observed in Pb [18, 19]. We temporally assign these dip-hump features to quasiparticle coupling to certain collective bosonic excitations. To read out the bosonic mode energy, we calculated numerically the second derivative of the tunneling conductance [bottom panel, Fig. 3(a)]. By subtracting Δ ~11.7 meV from the two dips (peaks) at energies of + (-) 21.7 meV and + (-) 34.0 meV, we obtained the energies of two bosonic modes $\Omega_1$ = 10.0 meV and $\Omega_2$ = 22.3 meV, respectively.

The observation of the bosonic modes raises immediately a question whether and how they are linked to superconductivity. We then conducted an extensive variable temperature (4.6 K to 30.2 K) STS experiment to identify its nature. The result for K-coated 2-UC films is shown in Fig. 3(b). With increasing temperature, the dip-hump features and the superconducting coherent peaks tend to degrade simultaneously and disappear eventually, suggesting that they are intertwined. For a more quantitative analysis, we measured the mode energy $\Omega$ from the STS spectra of superconducting films, including bare 1-UC films and K-coated 2−4 UC films at various K coverage (Fig. S2), and summarized the data in Fig. 3(c). Depending on the film thickness and K coverage, t4he superconducting gap changes significantly from 6.5 meV to 19 meV. The large fluctuation in superconducting gap for given thickness and K coverage is due probably to sample quality change in different runs of experiment, which is not a central problem of this study and will not be discussed here. Regardless of this variation, however, we see that the energy distribution of the bosonic modes collapses basically into two distinct groups, the first centered at 11.0 meV ($\Omega_1 = 11.0 \pm 2.1$ meV) and the second at 21.5 meV ($\Omega_2 = 21.5 \pm 4.5$ meV). This points out two characteristic energy scales for the bosonic modes in the system, which we emphasize are basically independent of doping level (K coverage) and film thickness.

We note that there is a striking consistency between the bosonic modes at $\Omega_1 = 11.0$ meV and $\Omega_2 = 21.5$ meV in the aforementioned superconducting films and the phonon frequencies in bulk materials. According to previous neutron scattering and Raman scattering measurements [27-29], the bulk FeSe exhibits $E_g$ (Se) phonons at frequency of 12−13.1 meV and $A_{1g}$ (Se) phonons at frequency ~19.8 meV, while STO has a $TO_2$ phonon with frequency of 21.7 meV. Albeit subtle change in frequency at surface/interface and probably further change caused by K adsorption, this agreement justifies our measurement and strongly suggests that the two bosonic modes are phonons. Therefore, the features in tunneling spectra shown in Fig. 3 correspond to the quasiparticle couplings to the phonons at ~11.0 meV and ~21.5 meV. The large superconducting gaps observed in Fig. 2 should result from strong *e-ph* coupling, implying that the pairing mechanism here is rather conventional. This interpretation is consistent with the U-shaped gap observed in the FeSe films on STO. In sharp contrast, in ultra-thin films of FeSe grown on and weakly bonded to graphene that we previously studied [12, 30], the superconducting gap decreases with decreasing film thickness and is vanishing for 1-UC films. Although a bosonic mode was also observed, the energy is smaller and only in a level of 2.7−4 meV [25]. These differences point out a special role of the STO substrate in the high $T_C$ superconductivity in ultra-thin films of FeSe we studied here.

The template effect in 1-UC FeSe films on STO was recently investigated theoretically [31]. According to the first-principles calculations, the STO substrate stabilizes the 1-UC FeSe films to a nearly square arrangement so as to prevent the films from undergoing a shear-type structure transition as the case in bulk. As a result, two *e-ph* coupling channels are opened and enhanced by the interface effect, which leads to an *e-ph* coupling constant $\lambda = 1.6$ which is ten times of that ($\lambda = 0.16$) of bulk FeSe [32]. The phonon frequencies involved in the coupling discovered from the calculations are 10 meV and 20 meV [31], which are, again, in excellent agreement with $\Omega_1$ ~11.0 meV and $\Omega_2$ ~21.5 meV for 1−4 UC films observed experimentally here.

To further understand the special role of STO substrate, we measured the in-plane lattice constant of 1−3 UC films on STO using STM. The atomically resolved STM images in Fig. 4 show that the films all exhibit a square lattice and that local distortion becomes weaker with increasing thickness. However, their fast Fourier transformation (FFT) in the insets reveals consistent in-plane lattice constant ~0.39 nm, which means that all the films are fully strained with STO (001) substrate. The lattice-mismatch associated strain may be partially released vertically due to the weak van der Waals interaction. Nevertheless, the data in Fig. 4 indicate that the template effect from the STO substrate is indeed significant and can persist at least up to 3-UC.

The interface promoted *e-ph* coupling is further supported by recent ARPES studies [5, 33]. The study with femtosecond time resolution not only identifies a group of phonons with a frequency consistent with the phonon mode $\Omega_2$, but also reveals that the phonons become soft at the interface[33]. The phonon softening at interface has long known as an effective way to enhance *e-ph* coupling strength and to raise $T_C$, as demonstrated in the early study of conventional superconductor multilayers [34]. Coupling to high frequency phonons is necessary to account for the high $T_C$ in the ultrathin FeSe films on STO [31], indeed, oxygen optical phonons at 100 meV was disclosed experimentally [5]. The resulted coupling to this oxygen optical phonons from calculations is sufficiently strong to account for $T_C$ ~40−50 K, without the need for any additional unconventional pairing mechanism [35]. Although quantitative agreement in $T_C$ between the experiments and theories has not been reached at this stage, all available data can qualitatively be understood consistently within the *e-ph* coupling scenario.

The interface enhanced *e-ph* coupling and hence high $T_C$ superconductivity is also revealed in 1-UC FeTe$_{1-x}$Se$_x$ films on STO. In this case, the U-shaped superconducting gaps ranges from 13 meV to ~16.5 meV [36], nearly ten times the gap value (~1.7 meV) of the optimally doped bulk FeTe$_{0.6}$Se$_{0.4}$ single crystal [37]. Several groups of phonons with frequency of ~10 meV, ~20 meV and ~25 meV are observed on 1-UC FeTe$_{1-x}$Se$_x$ films (0.1 ≤ x ≤ 0.6) on STO (Fig. S3). These phonon modes are consistent with $E_g$(Te/Se), $A_{1g}$(Te/Se)/TO$_2$(STO) and $B_{1g}$(Fe) modes, respectively, observed in previous Raman and neutron scattering studies [29, 38, 39]. The spin resonance mode at ~6 meV [40, 41] is not observed. The above result indicates that interface enhanced *e-ph* coupling is a rather general approach for raising superconductivity temperature.

Finally, we briefly comment on the interpretation of the dome-shaped superconducting phase diagram in terms of interface enhanced antiferromagnetic fluctuation [16]. Within the antiferromagnetic fluctuation scenario, different doping level should alter spin-fluctuation by promoting different ordered phases and change the spin resonance energy [24, 42]. However, what we observed is that the mode energies are independent of superconducting gap and doping level, as shown in Fig. 3(c) and Fig. S2. On the other hand, because of the change of screening [43], the amplitude of *e-ph* coupling can be doping dependent. And, quasiparticle scattering from K atoms/clusters can also lead to suppression of superconductivity (Fig. S4), and thus contribute to the dome-shaped superconducting phase diagram.

In summary, our study demonstrates that 2−4 UC FeSe films on STO all exhibit superconducting gap larger than 11 meV under optimal surface doping of K atoms/clusters and interface enhanced *e-ph* coupling is the most plausible mechanism for the high temperature superconductivity in ultrathin FeSe films on STO. In terms of the similar doping mechanism associated with the built-in heterostructure in cuprates and Fe-based superconductors, our study hints that the superconducting gap in the $CuO_2$ and FeAs(Se) planes might be similar to that of FeSe films on STO and nodeless too and that the high $T_C$ in cuprates and Fe-based superconductors can be understood within the *e-ph* coupling BCS theory. Under this context, direct measurement of those superconducting planes becomes exceedingly crucial for understanding high $T_C$ superconductivity. Motivated by this work, we propose to fabricate heterostructure where 2−3 UC FeSe sandwiched between STO on both sides or STO on one side and $A_xFe_{2-y}Se_2$ (A=K, Rb, Cs, Tl/K) on the other side to achieve higher $T_C$.

This work is supported by NSFC (11321091, 91121004) and MOST of China (2015CB921000).

C.J. Tang and C. Liu contributed equally to this work.

[*] liliwang@mail.tsinghua.edu.cn

[†] qkxue@mail.tsinghua.edu.cn

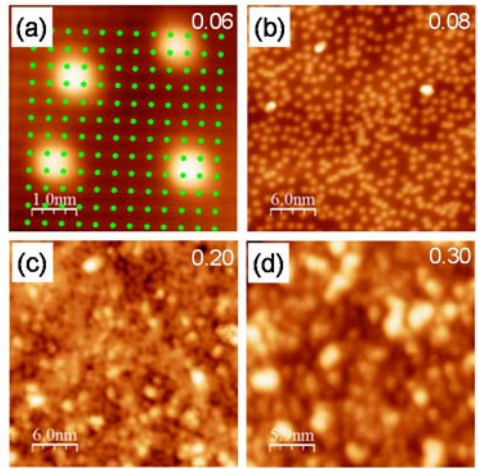

FIG. 1 (color). Topographic images of the surface with K adsorption at various coverage. (a) 0.06 ML ($V = 500$ mV, $I = 50$ pA), (b) 0.08 ML, (c) 0.20 ML and (d) 0.30 ML ($V = 1$V, $I = 30$ pA). The green dots in (a) show the position of Se atoms.

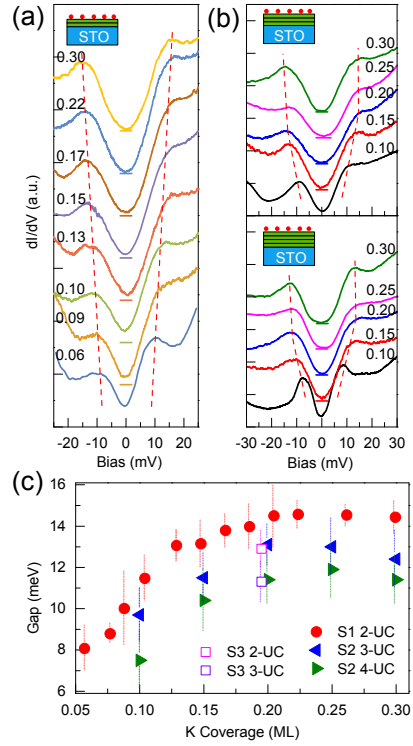

FIG. 2 (color). Typical *dI/dV* curves taken on the 2-UC (a) and 3−4 UC (b) FeSe films at various K coverage ($V$ = 30 mV, $I$ = 100 pA). The horizontal bars indicate zero conductance position of each curve. The dashes are guide for eyes, showing the change of coherence peaks. (c) The dependence of the superconducting gaps on K coverage.

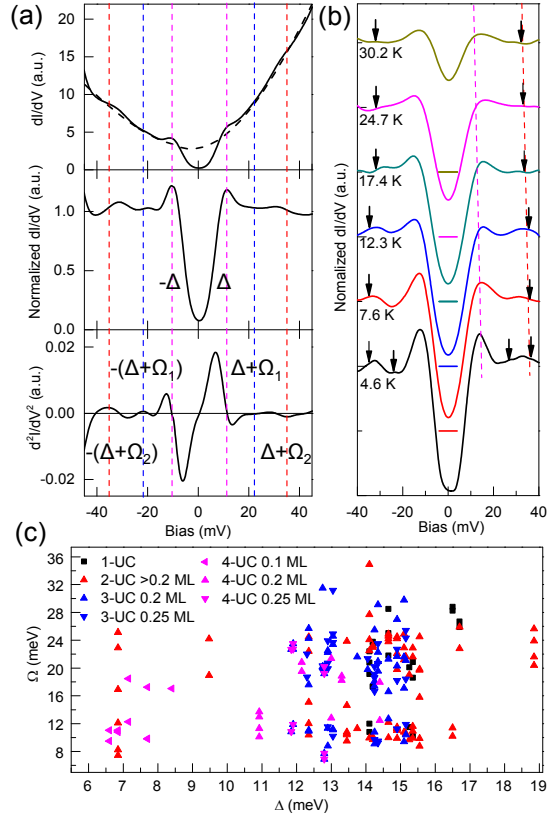

FIG. 3 (color). (a) Black curves show raw *dI/dV* (top panel), normalized *dI/dV* (middle panel), and *d²I/dV²* (bottom panel) spectra on 3-UC films at the K coverage of 0.20 ML. The normalization was performed by dividing the raw *dI/dV* spectrum by its background, which was extracted from a cubic fit to the conductance for $|V| > 20$ mV (the dashed line in the top panel). The pink, blue and red dashes show the approximate energy positions of $\pm \Delta$, $\pm (\Delta + \Omega_1)$ and $\pm (\Delta + \Omega_2)$, respectively. (b) Normalized *dI/dV* spectra at temperatures ranging from 4.6 K to 30.2 K taken on 2-UC FeSe films at the K coverage of 0.20 ML. The horizontal bars indicate zero conductance position of each curve. The pink dash and the red dash are parallel and show the synchronous change in coherence peak and feature of phonon $\Omega_2$. (c) The distribution of the phonon energy $\Omega$ as a function of the superconducting gap magnitude $\Delta$.

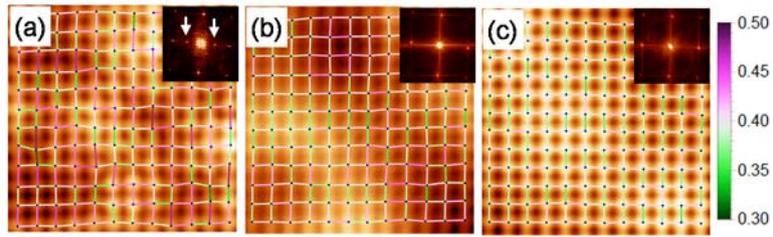

FIG. 4 (color). Atomically resolved images of 1-UC (a), 2-UC (b) and 3-UC FeSe (c) films acquired on one single sample ($V$ = 100 mV, $I$ = 100 pA, 4.8 nm × 4.8 nm). Local maxima (black dots) are used as approximate position of Se atoms, the distance between adjacent atoms are manifested by the colored segments. The insets are FFT images based on an area of 30 nm × 30 nm. The arrows in (a) label the spots corresponding to the 2 × 1 electronic structure, which is unique feature of 1-UC films.

## Supplementary Material for:

## Interface enhanced electron-phonon coupling and high temperature superconductivity in potassium-coated ultra-thin FeSe films on SrTiO$_3$

Chenjia Tang,[1] Chong Liu,[1] Guanyu Zhou,[1] Fangsen Li,[1] Ding Zhang,[1] Zheng Li,[1] Canli Song,[1,2] Shuaihua Ji,[1,2] Ke He,[1,2] Xi Chen,[1,2] Lili Wang,[1,2,*] Xucun Ma,[1,2] and Qi-Kun Xue[1,2,†]

[1] *State Key Laboratory of Low-Dimensional Quantum Physics, Department of Physics, Tsinghua University, Beijing 100084, People's Republic of China*

[2] *Collaborative Innovation Center of Quantum Matter, Beijing 100084, People's Republic of China*

**MATERIALS AND METHODS**

The experiments were conducted in an ultrahigh vacuum (UHV) low temperature (4.6 K) STM system equipped with a molecular beam epitaxy (MBE) chamber for film growth (Createc Fischer & Co. GmbH). The base pressure of the system is better than $2\times10^{-10}$ Torr. For FeSe film growth, high purity Fe (99.995%) and Se (99.999%) at a nominal Se/Fe beam flux ratio of ~10 were co-deposited onto the STO(001) substrate held at 400°C. To achieve superconductivity, the samples were subsequently annealed at 470 °C for several hours. Potassium (K) atoms were deposited onto the FeSe films cooled down below 100 K by liquid nitrogen. After K deposition, the samples were immediately transferred to the STM stage cooled at 4.6 K for STM measurements. At temperature below 100 K, the K atoms mainly adsorb on the topmost surface of the FeSe films.

We have prepared FeSe films with nominal thickness of 1.5 UC, 2.5 UC and 3.5 UC. Due to the layer-by-layer growth mode, two adjacent thicknesses are observed on the surface, for example, 1-UC and 2-UC FeSe films coexist on the surface of the 1.5 UC sample. This situation allows us directly compare the properties of adjacent layers on the same sample. As shown in Fig. 2(c), the comparisons between 3-UC and 4-UC films on the 3.5 UC sample, and 2-UC and 3-UC on the 2.5 UC sample show consistent results.

In all STM/STS measurements, a polycrystalline PtIr tip was used. The STS was acquired by using lock-in technique with a bias modulation of 0.5 mV at 437 Hz and set point of 30 mV, 100 pA.

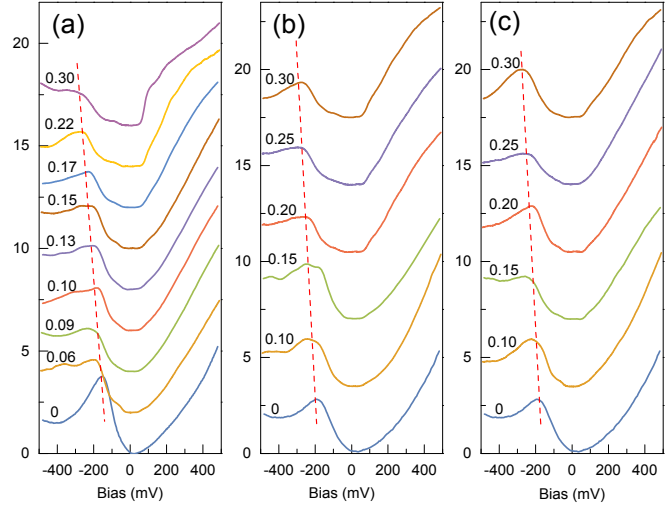

FIG. S1 (color). K doping effect in 2−4 UC FeSe films. Tunneling spectra ($V$ = 500 mV, $I$ = 100 pA) of (a) 2-UC, (b) 3-UC and (c) 4-UC at K coverage from 0 to 0.30 ML. Electron doping effect with K adsorption can be seen by a systematic downward shift (red dashes) of valence band with increasing K coverage. The doping effect becomes saturated above 0.20 ML where the adsorbed K atoms start to form clusters.
15

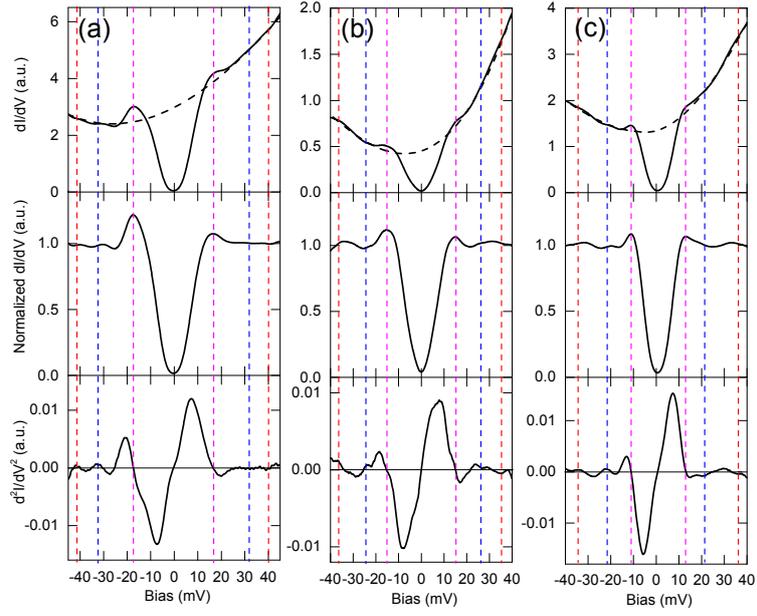

FIG. S2 (color). The phonon modes in ultra-thin FeSe films at various K coverage. (a) Bare 1-UC FeSe film, (b) 2-UC FeSe film with 0.30 ML K and (c) 4-UC FeSe film with 0.30 ML K. Black curves show the raw *dI/dV* (top panel), normalized *dI/dV* (middle panel), and $d^2I/dV^2$ (bottom panel) spectra, respectively. The normalization was performed by dividing the raw d*I*/d*V* spectrum by its background, which was extracted from a cubic fit to the conductance for $|V| > 20$ mV (the dashed line in the top panel). The pink, blue and red dashes show the approximate energy positions of $\pm \Delta$, $\pm (\Delta + \Omega_1)$ and $\pm (\Delta + \Omega_2)$, respectively.

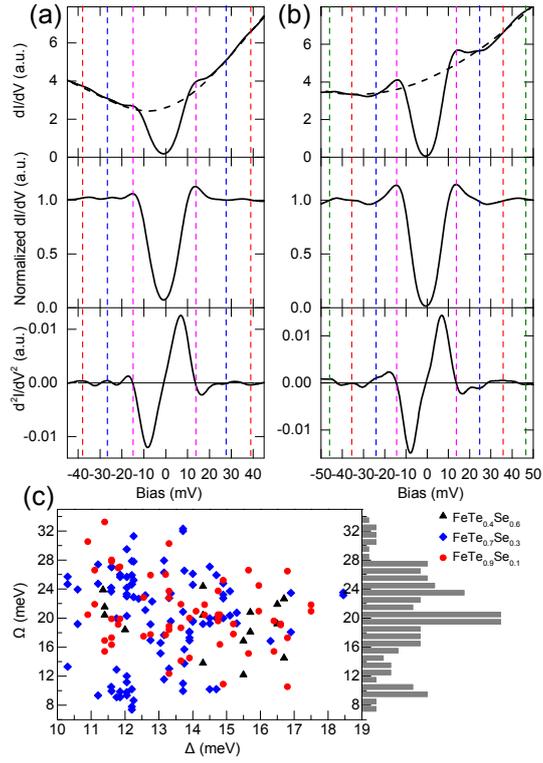

FIG. S3 (color).    The phonon modes in 1-UC FeTe$_{1-x}$Se$_x$ films on STO. (a) FeTe$_{0.4}$Se$_{0.6}$ and (b) FeTe$_{0.7}$Se$_{0.3}$. Black curves show the raw d$I$/d$V$ (top panel), normalized d$I$/d$V$ (middle panel), and d$^2I$/d$V^2$ (bottom panel) spectra, respectively. The normalization was performed by dividing the raw d$I$/d$V$ spectrum by its background, which was extracted from a cubic fit to the conductance for $|V| > 20$ mV (the dashed line in the top panel). The pink, blue, red and green dashes show the approximate energy positions of $\pm \Delta$, $\pm (\Delta + \Omega_1)$, $\pm (\Delta + \Omega_2)$ and $\pm (\Delta + \Omega_3)$, respectively. (c) The distribution of phonon energy $\Omega$ as a function of the superconducting gap magnitude $\Delta$.

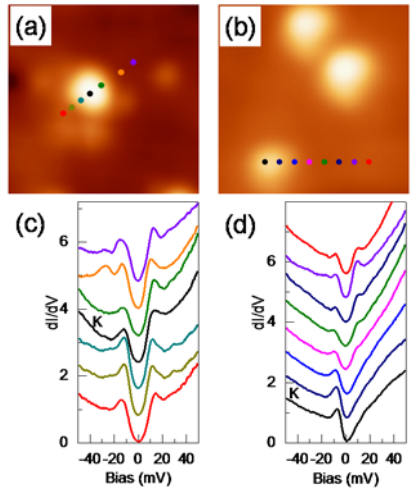

FIG. S4 (color). (a) and (b) Topographic images of the 2-UC FeSe films with K adatoms. (c) and (d) Spectra taken on the dots shown in (a) and (b), respectively. On K atoms, the coherence peaks are suppressed (c) and even vanishing (d).